\documentclass[11pt,twocolumn]{article}
\usepackage[dvips]{graphicx}

\begin{document}

\title {Self Consistent Field Theory of
Twist Grain Boundaries in Block Copolymers}
\author{%
Daniel Duque$^1$, M. Schick$^2$\\
$^1$Departamento de F{\'{\i}}sica
Te\'orica de la Materia Condensada,
\\ Universidad Aut\'onoma de Madrid,
E-28049 Madrid, Spain\\
$^2$Department of Physics, Box 351560, University of
Washington\\
Seattle WA, 98195-1560 USA\\ }

\maketitle

\begin{abstract}

We apply self consistent field theory to twist grain boundaries of block
copolymer melts. The distribution of monomers throughout the grain boundary
is obtained as well as the grain boundary free energy per unit area as a
function of twist angle. We define an intermaterial dividing surface in
order to compare it with minimal surfaces which have been proposed. Our
calculation shows that the dividing surface is not a minimal one, but the
linear stack of dislocations seems to be a better representation of it for
most angles than is Scherck's first surface.

\end{abstract}

\section{Introduction}

Bulk equilibrium properties of diblock copolymer melts are
relatively well understood\cite{hamley}. Incompatibility  of the
monomers comprising the two blocks drives the system toward
ordered structures in which the number of contacts between
dissimilar monomers is reduced, subject to various constraints.
These ordered phases, of which the simplest is lamellar,  are
thermodynamically stable below some order-disorder transition
temperature.

When the system is taken below this temperature, the lamellar phase is
nucleated typically in distinct grains which differ, one from the other, by
the orientation of the lamellae within them. The interface between lamellae
of different grains constitutes a grain boundary, which can be considered an
equilibrium structure arising from a constraint that imposes the different
orientations of the lamellae of the two grains.

Because the lamellae of block copolymers lack any internal order, their
grain boundaries are simpler than those between grains of crystalline
solids. While the latter are combinations of five different independent
boundaries, the former can be decomposed into only two independent ones.  In
the kink grain boundary, the normals perpendicular to the lamellae of the
two grains define a plane which is \emph{perpendicular} to the plane of the
boundary. Kink grain boundaries have been studied recently both
experimentally \cite{gidokgb,hashimoto1,nishikawa} and theoretically
\cite{nas,matsen,tas}. In the twist grain boundary (TGB), the plane defined
by the normals is \emph{parallel} to the plane of the boundary. The angle,
$\alpha$, between the normals defines the twist angle. The geometry is shown
in Fig. 1 where we fix the convention we use throughout: $x$ is the
direction perpendicular to the grain boundary and $y$ and $z$ are in the
plane of it; normals to the lamellae of the two grains are at angles
$\pm\alpha/2$ with respect to the $y$ axis. Twist grain boundaries have been
the subject of a thorough experimental study\cite{gidotgb}, but have
received less theoretical attention.

\begin{figure}
\fbox{%
\includegraphics[width=75mm]{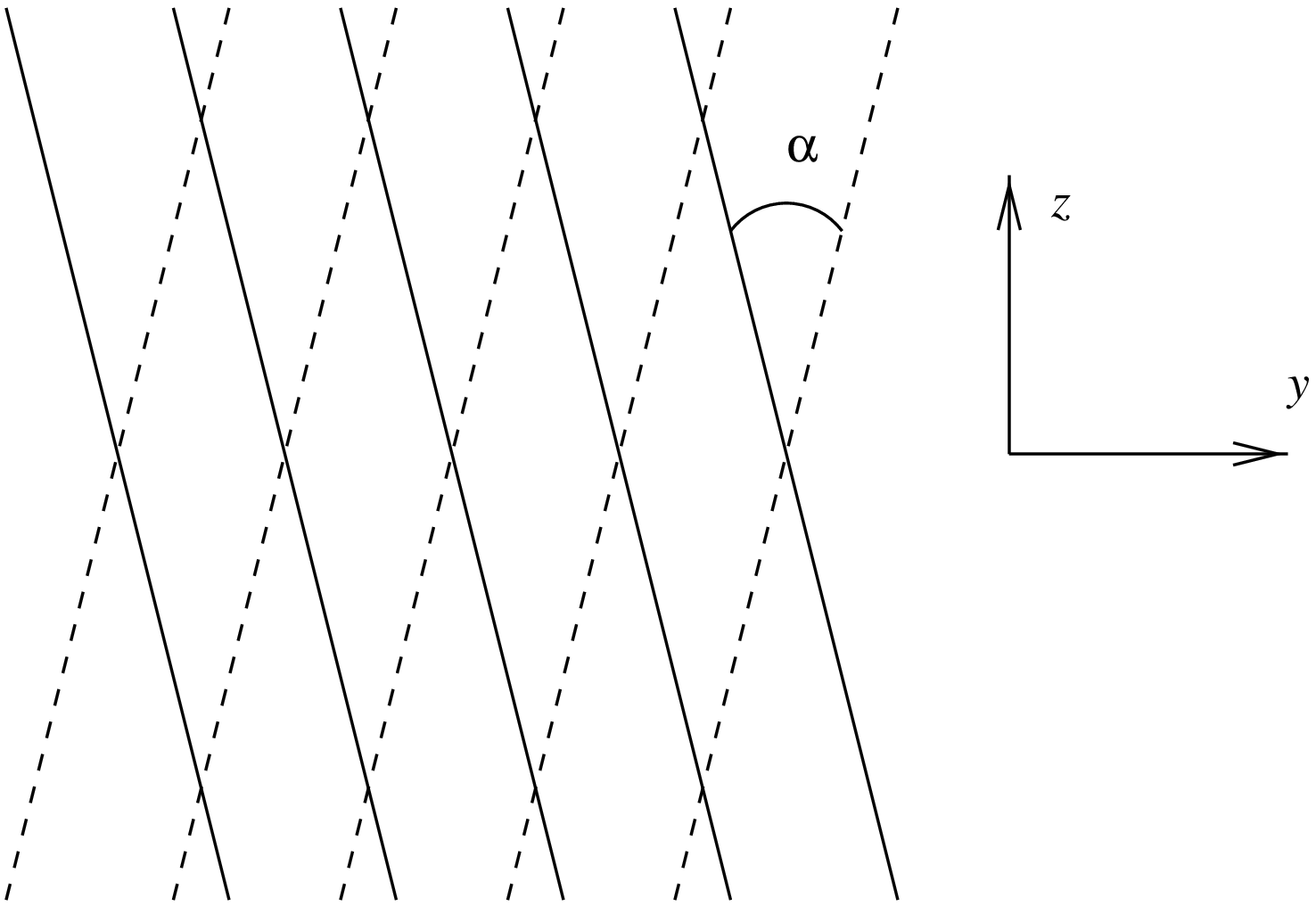}
}
\caption{Schematic of a twist grain boundary in which one looks
down the $x$ axis. As $x\rightarrow\infty$, the IMDS approach that of the
lamellar bulk, shown in dashed lines, while as  $x\rightarrow -\infty$, the
IMDS approach that of the lamellar bulk shown in solid lines. Normals to the
planes of the bulk lamellae make angles of $\pm\alpha/2$ with respect to the
$y$ axis.}
\end{figure}

Two different kinds of twist grain boundaries have been observed. The
simplest consists of a stack of planes. In each plane, the orientation of
lamellae differs slightly from those in planes above and below it. The
structure is periodic along the stacking direction.  It is observed  for
small twist angles ($\alpha<15^\circ$) only. A heuristic model for this
boundary was proposed by Gido \textit{et al.} \cite{gidotgb} and checked
against their experimental data, with good results.

The other structure, observed for all angles,  is quite different: it is
\textit{doubly} periodic. A surface which displays this double periodicity
is Scherck's first surface, one which is characterized by zero mean
curvature everywhere. It is shown in Fig. 2 for a twist angle $\alpha= 0.5$
radians. This minimal surface was first proposed as a model of the TGB by
Thomas \textit{et al.}\cite{thomas}. The reasoning is as follows. Suppose
that one can ignore the components of the system, the block copolymers and
the constituent monomers, $A$ and $B$ of the two blocks, and focus instead
on the internal interfaces which divide the $A$-rich lamellae from the
$B$-rich ones. Suppose further that one can ignore the finite thickness of
these internal interfaces and approximate them by a suitably defined
surface, the intermaterial dividing surface, (IMDS). The area of this
surface is extensive, \textit{i.e.} proportional to the volume of the
system. Under these assumptions, the bulk free energy of the system should
be expressable as the energy of this surface. Minimization of this free
energy leads to the surface with minimum area subject to the constraint that
it separate regions of certain volume. This is a surface of constant mean
curvature. In particular for a symmetric system in which the volumes are
equal, and that is the case for a diblock with equal volumes of $A$ and $B$
monomers, the constant value of the mean curvature is zero. Such surfaces
are called minimal surfaces. The choice of the appropriate minimal surface
depends upon boundary conditions and expected symmetries. Scherck's first
surface\cite{nitsche} was chosen in Ref \cite{gidotgb} for comparison with
experimental results because it connects two sets of parallel planes, the
normals of which form an angle $\alpha$. The explicit expression for the
surface is quite simple:
\begin{equation}
\label{scherck}
\exp\left[\frac{2\pi}{D} x \sin\alpha\right]=
  \frac{\cos\left[\frac{2\pi}{D} \{\cos(\alpha/2)y+\sin(\alpha/2)z\}\right]}
       {\cos\left[\frac{2\pi}{D} \{\cos(\alpha/2)y-\sin(\alpha/2)z\}\right]},
\end{equation}
The equation defining the surface pertains only in the regions for which the
right hand side is positive. This condition defines a chessboard-like domain
in the $y$-$z$ plane (as a consequence of which the surface is sometimes
referred to as ``Scherck's doubly periodic surface''). Far from the grain
boundary, whose center is at $x=0$, the nature of the surface is easily
inferred from Eq. \ref{scherck}: when $x\to-\infty$ the r.h.s. must vanish,
and this defines a set of parallel planes with unit normal
$(0,\cos(\alpha/2),\sin(\alpha/2))$ and spacing $D/2$. The period of the
IMDS is half that of the lamellar phase, $D$, hence our convention.
Similarly, when  $x\to\infty$ the r.h.s. must diverge, and this defines a
set of parallel planes with unit normal $(0,\cos(\alpha/2),-\sin(\alpha/2))$
and spacing $D/2$.
\begin{figure}

\includegraphics[width=75mm]{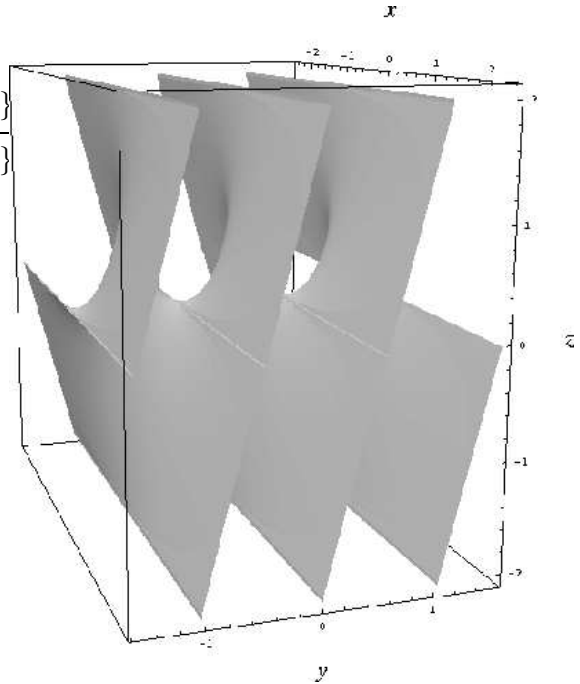}
\caption{Scherck's first surface is shown for a twist angle of
$\alpha=0.5$ and a scale $D=1$.}
\end{figure}

A different description of the grain boundary, due to Renn and Lubensky
\cite{renn}, is that of a linear stack of dislocations (LSD). The model of a
single dislocation is again a minimal surface, the helicoid. The whole
structure arises from the stacking of
infinitely many dislocations along a line.
In this case the dislocations have vorticity along the $y$
axis with a pitch $D$, and are stacked along the $z$ axis with a separation
$D/(2\sin(\alpha/2))$. It was later shown \cite{k&l} that this approach is
actually equivalent to the description employing Scherck's surface, up to a
$\cos(\alpha/2)$ dilation of the $x$ axis, the LSD being more ``compressed''
than Scherck's surface.

The above approaches are valuable in providing simple models of the grain
boundary which can be compared with experiment. They suffer, however, from
the approximations which are inherent in the approach. Most importantly in
the system of block copolymers, they ignore the physical constraints of
incompressibility which causes the chains to stretch in order to fill the
available volume. As a consequence, the IMDS is not a surface of constant
mean curvature, a point made compellingly by Matsen and Bates\cite{mandb},
and confirmed in experiment\cite{hashimoto}.

Therefore we have studied the twist grain boundary in block copolymer
systems using the self-consistent field theory in Fourier space which has
been successful in describing lyotropic phases of block copolymer
melts\cite{mands}. We follow the approach of Matsen\cite{matsen} who adapted
it to kink grain boundaries. First we will introduce the particular
implementation of the theory which is useful in this case. We then  present
results for the TGB obtained within this framework and compare them with
Scherck's first surface and the LSD. We conclude with a brief discussion.

\section{Theory}

We consider an incompressible melt of $n$ AB diblock copolymers each
composed of $N$ segments of volume $1/\rho_0$; the volume of the system is
$V=nN/\rho_0$. The polymers are modeled as Gaussian random walks with
statistical step length $a$. The natural length scale of the problem is the
end to end mean distance, $a N^{1/2}$. To describe the incompatibility
between A and B monomers the standard Flory-Huggins parameter, $\chi$, is
introduced; the product $\chi N$ sets the units of energy and temperature.

\begin{figure}

\includegraphics[width=75mm]{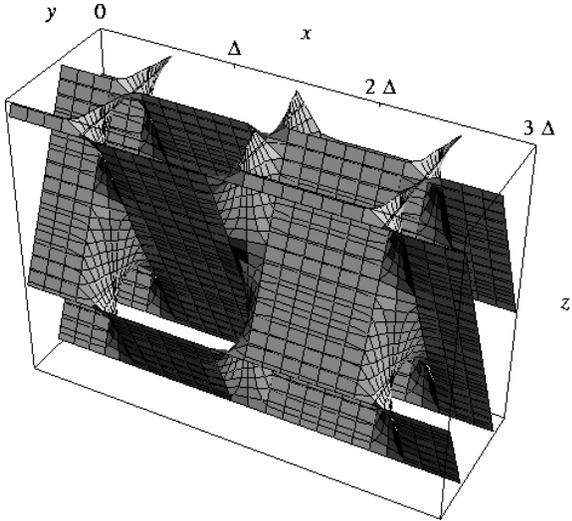}
\caption{%
Schematic of the system, periodic in all three
directions, on which we perform our calculations. The original system of
interest is recovered on letting $\Delta$ increase without limit.}
\end{figure}

We utilize the SCFT method expressed in Fourier space as in Ref.
\cite{mands}, a method suited to the study of periodic phases. A system
containing a twist grain boundary, however,  is only periodic in the
coordinates $y$ and $z$, but not in $x$. In order to circumvent this
difficulty, we adopt the same strategy employed by  Matsen in his study of
the kink grain boundary\cite{matsen}: to express the desired system as one
periodic in all three directions in the limit in which one period becomes
infinite. We therefore consider the system shown in Fig. 3 which, in
addition to being periodic in $y$ and $z$ is periodic in $x$ with period
$\Delta.$ We  impose a reflection symmetry around $x=\Delta/2$,
\textit{i.e.}, $(x,y)\to(\Delta-x,y)$. The desired grain boundary free
energy per unit area, $\Gamma$, of the original system is obtained from the
free energy, $F(\Delta)$, of our system according to
\begin{equation}
\Gamma=\lim_{\Delta\to\infty}[F(\Delta)-F_{\mathrm{b}}]\frac{\Delta}{V},
\end{equation}
where the bulk free energy
$F_{\mathrm{b}}= \lim_{\Delta\to\infty}F(\Delta)$, and the
area of the grain boundary is $V/\Delta$. As the natural unit of area per
polymer is $V/naN^{1/2}$, the natural unit of the grain boundary free energy
is $k_{\mathrm{B}}TnaN^{1/2}/V$. Thus the boundary free energy can be written
\begin{equation}
\label{tension} \gamma\equiv
\frac{V}{k_{\mathrm{B}}T n aN^{1/2}}\Gamma=
\lim_{\Delta\to\infty}
 \frac{F(\Delta)-F_{\mathrm{b}}}{k_{\mathrm{B}}Tn}
	\frac{\Delta}{aN^{1/2}}
\end{equation}

As the system is now periodic in all directions, we can expand all functions
of position into a complete, orthonormal, set of eigenfunctions, $f_{lmn}$,
of the Laplacian operator\cite{mands}, eigenfunctions which explicitly
express the symmetries of the system. It is clear that it is invariant under
a rotation of $\pi$ about the $x$ axis, {\it i.e.} under
$(x,y,z)\to(x,-y,-z)$. It is also invariant under rotations of $\pi$ about
the $y$ and $z$ axes as well. Therefore the system is invariant with respect
to the change in sign of any two of the coordinates. These considerations,
together with the imposed reflection symmetry around $x=\Delta/2$, lead to
the choice of functions
\begin{equation}
\label{basis}
\begin{array}{l}
f_{l m n}(x,y,z)= \\
=\left\{
  \begin{array}{l}
       c_l c_m c_n \cos(l k_x x) \cos(m k_y y) \cos(n k_z z), \\
       \qquad\qquad         \textrm{if $l$ is even}    \\
       c_l c_m c_n \sin(l k_x x) \sin(m k_y y) \sin(n k_z z), \\
       \qquad\qquad         \textrm{if $l$ is odd},    \\
  \end{array}
\right.
\end{array}
\end{equation}
where $k_x=\pi/\Delta$, $k_y=2\pi\cos(\alpha/2)/D$ and
$k_z=2\pi\sin(\alpha/2)/D$. Again $D$ is the bulk lamellar period.

It should  be clear from Fig. 1 that the natural coordinates in which to
express the periodicity of the system are  $(y/D)\cos(\alpha/2)\pm
(z/D)\sin(\alpha/2)$. Indeed Scherck's surface, Eq. \ref{scherck}, is
expressed in them. Translating an expansion in those coordinates into an
expansion in $x$, $y$, and $z$, one sees that the parity of $m$ and $n$
above must be the same. Finally the $c_l$ are  determined by orthonormality:
\begin{equation}
\frac{1}{V}\int f_{lmn}(\mathbf{r})f_{l'm'n'}(\mathbf{r})d\mathbf{r}=
\delta_{ll'}\delta_{mm'}\delta_{nn'}.
\end{equation}
Thus $c_0=1$ and $c_i=\sqrt{2}$ for $i>0$.
This completes the specification of the basis functions.

One might think it necessary to impose an additional invariance: that the
calculated free energy of the grain boundary with twist angle
$\alpha\leq\pi/2$ be identical to that with angle $\pi-\alpha$ because,
after an interchange of $y$ and $z$, the one boundary is identical to the
other. However the grain boundary free energy we calculate already displays
the symmetry $\Gamma(\alpha)=\Gamma(\pi-\alpha)$ without further restriction
of the basis functions. This can be seen from the fact that under
$\alpha\rightarrow \pi-\alpha$, the wavevector components $k_y$ and $k_z$
interchange. Thus an interchange of the coordinates $y$ and $z$ and a
relabelling of the dummy indices $m$ and $n$ suffices to make the
expressions for the free energies of the two grain boundaries identical.

We expand all functions of position in terms of the above basis functions.
Of course the  infinite set must be truncated in a numerical calculation,
and  our computer resources impose a maximum slightly below $400$ functions.
The results presented below are obtained for the choice of $5$ values for
$m$, $5$ for $n$ and $15$ for $l$ as the results are more sensitive to the
number of components in the $x$ direction than in   $y$ or $z$. To be
consistent with this choice, we take the corresponding bulk lamellar phase
to be that obtained from $5$ Fourier components.

We have chosen $\chi N=15$, an intermediate value for which an intermaterial
dividing surface is well delineated, but not so sharp as to require  a large
number of basis functions to describe. A value of $\Delta\sim 5aN^{1/2}$ is
sufficient typically for the free energy to become insensitive to further
increases in this parameter. With our 375 basis functions, our results for
the free energy at $\alpha=\pi/2$
are accurate to within 1\%. Larger values of $\chi N$ would
require additional basis functions. Smaller values of $\chi N$, nearer the
order disorder transition temperature of $\chi N=10.49$, cause periodic
modulations of the dividing surface to appear which extend away from the
grain boundary. This behavior, similar to that reported  for the kink grain
boundary \cite{nas,matsen}, is likely to be strongly modified by fluctuation
effects which are absent in the SCFT.

\section{Results}

In Fig. 4 we show results for the twist angle $\alpha=0.4\approx
22.9^\circ$. We have plotted, for several values of $x$, contours of
constant order parameter, the difference between  the volume fractions of
the two monomers. In these gray scale plots, the maximum absolute value of
the order parameter is $0.88$. Fig. 4(a)  shows a slice at infinitely large
$x$, that is, a cross section through the bulk system. Figs. 4(b) and  (c)
show slices at the values of $x=0.2\ aN^{1/2}$, and $0$, that is, at the
grain boundary itself.

\begin{figure}

\begin{minipage}{85mm}
\centering
\includegraphics[width=20mm]{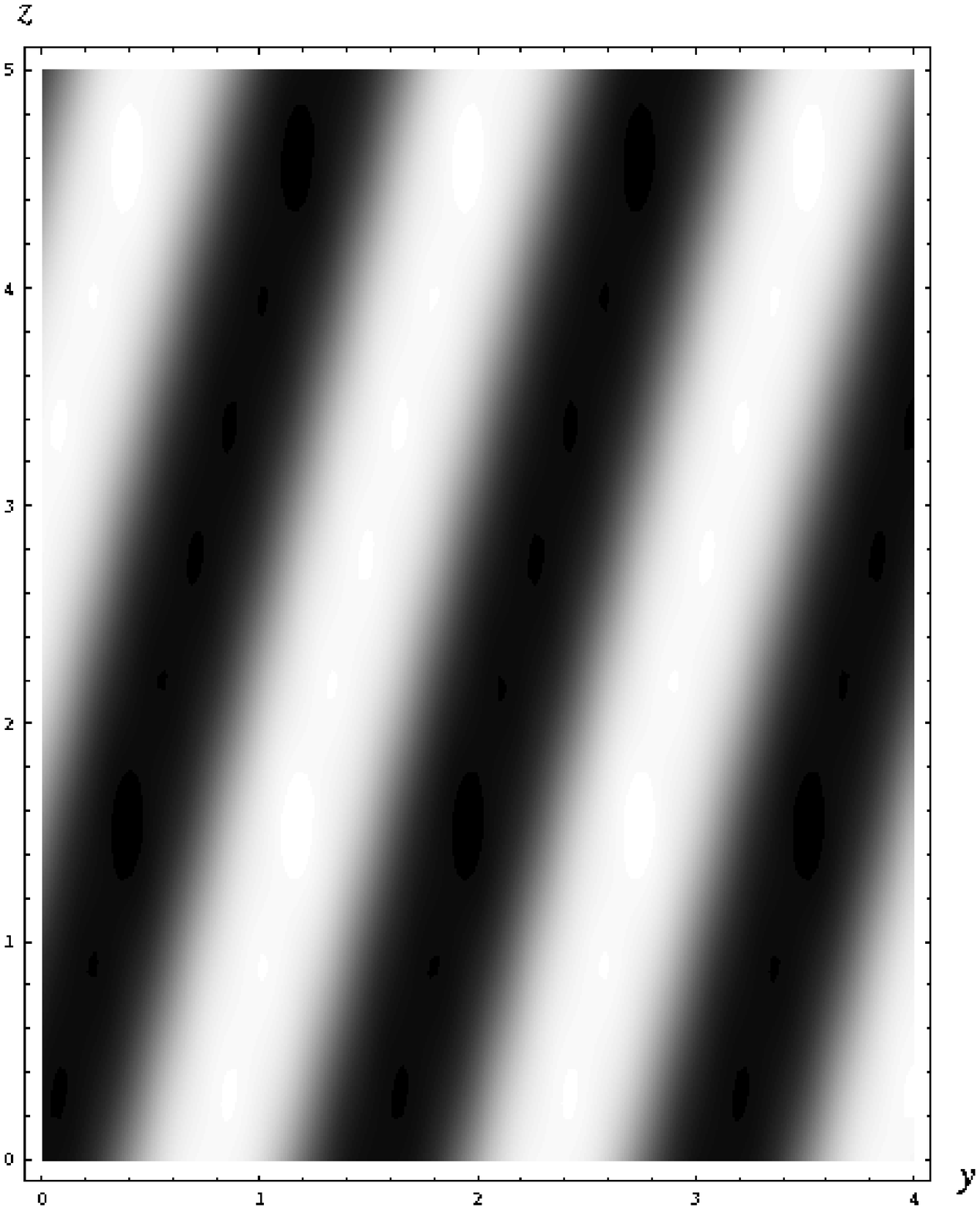}
\includegraphics[width=20mm]{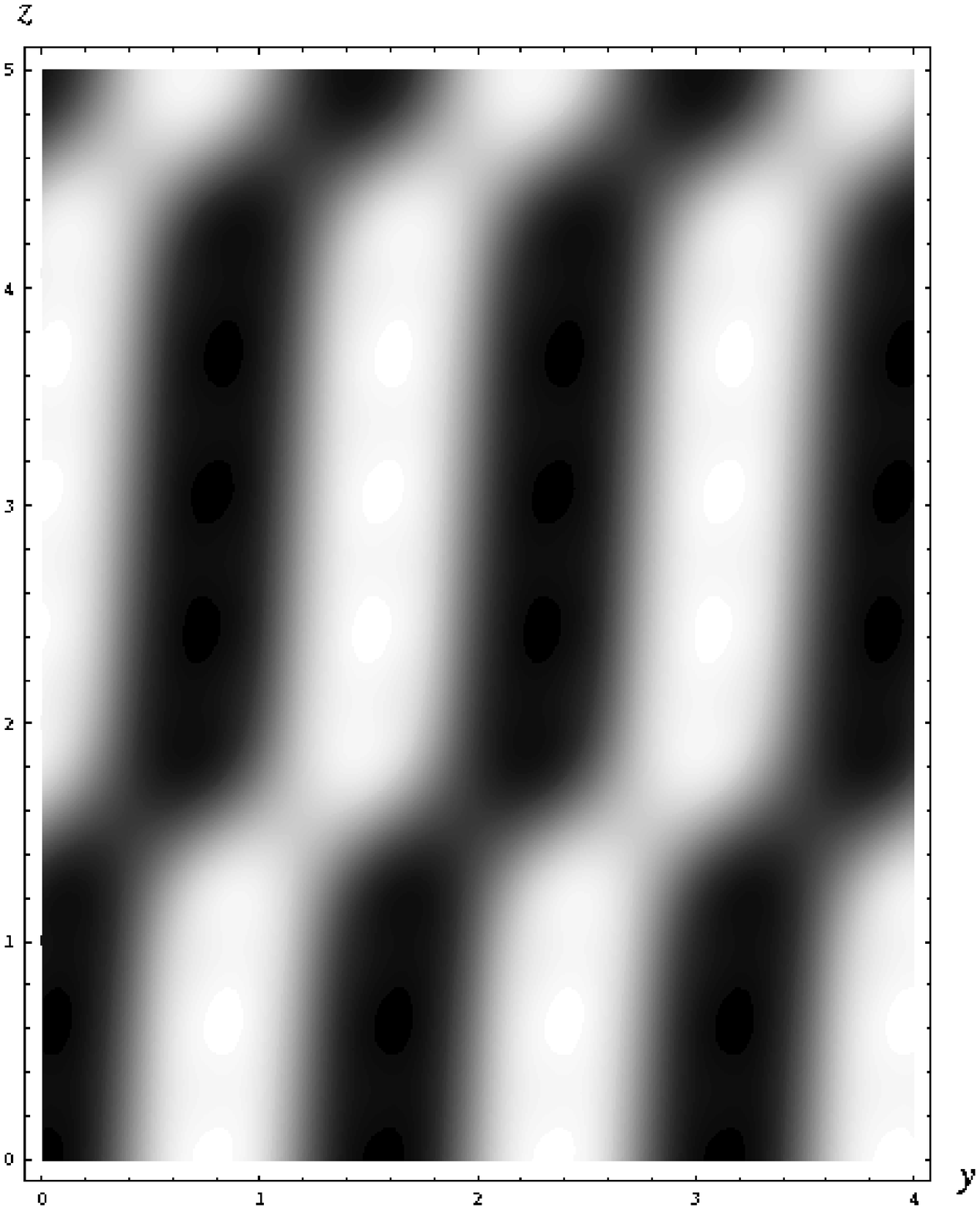}
\includegraphics[width=20mm]{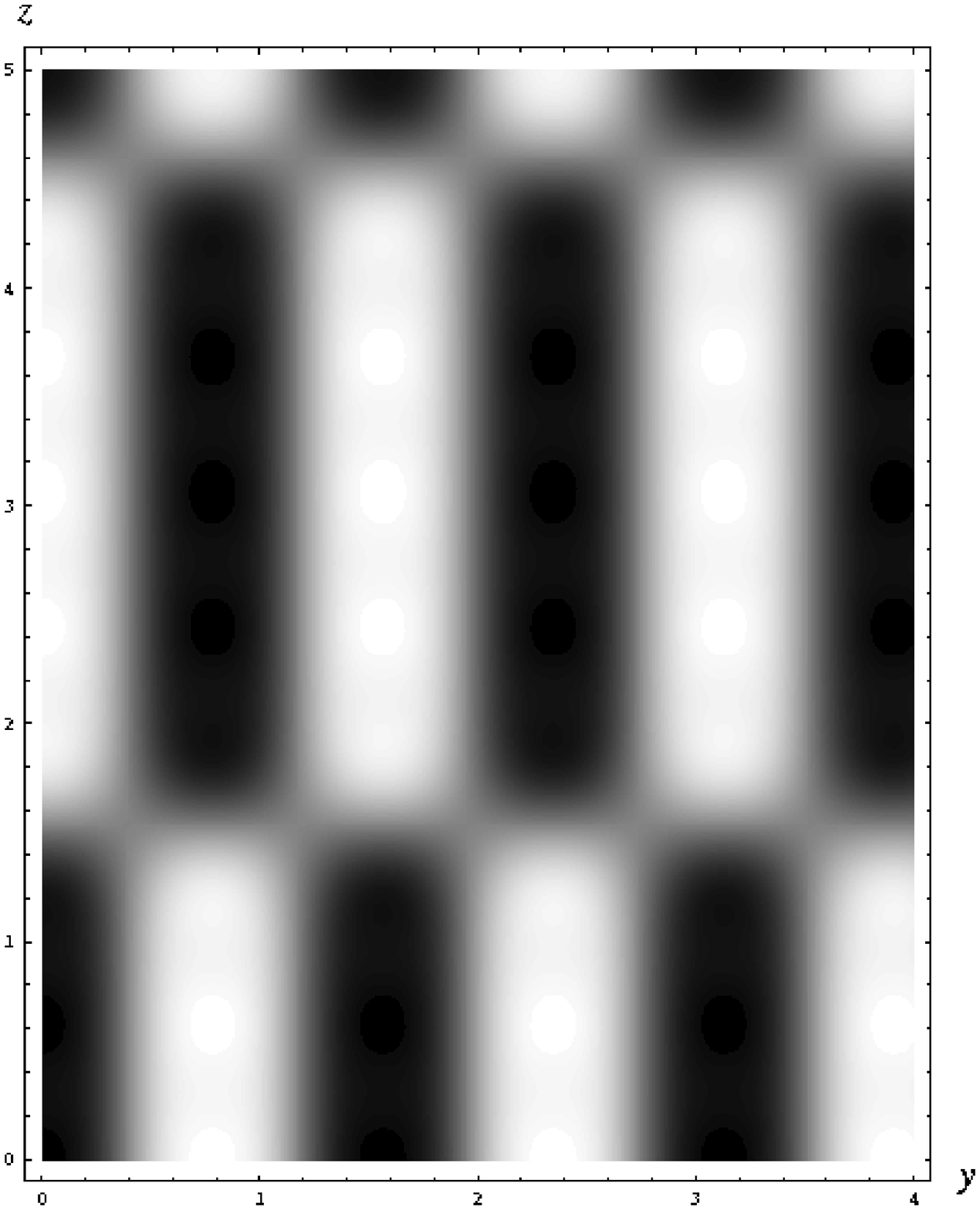}
\end{minipage}
\caption{%
Slices through the grain boundary showing the monomer
distribution at different distances: (a) In the limit of
$x\rightarrow\infty$, \textsl{i.e.} the bulk, (b) $x=0.2 aN^{1/2}$, (c)
$x=0$, \textsl{i.e.} at the grain boundary itself. In these gray scale
plots, the maximum absolute value of the order parameter is $0.88$.}
\end{figure}

We would like to compare the distribution of monomers obtained in our
solution  with Scherck's first surface, which is a model for the
intermaterial dividing surface.  One way to do this is to calculate within
our solution the value of the order parameter,
$\delta\phi(\mathbf{r})\equiv\phi_A(\mathbf{r})-1/2$,
at the points $\mathbf{r}_{\mathrm{S}}$ defined by
Scherk's surface.  The value of
$\delta\phi(\mathbf{r}_{\mathrm{S}})$ vanishes as
$x\rightarrow\pm\infty$ because Scherck's surface and the intermaterial
dividing surface of our solution, defined by
$\delta\phi(\mathbf{r}_{\mathrm{I}})=0$,
coincide in that limit. A convenient measure of the similarity of the two
dividing surfaces, therefore, can be defined by computing
\[
I_\mathrm{S}\equiv\int\ d\mathbf{r} \left[ \phi_A(\mathbf{r})-1/2
\right]^2\delta(\mathbf{r}-\mathbf{r}_{\mathrm{S}})
 \]
A measure $I_{\mathrm{LSD}}$ for the LSD can be defined in the same manner.

A second means to compare the surfaces is to calculate the volume of the
region which is enclosed between the two surfaces to be compared. This is
easy to implement by a Monte Carlo integration technique in which points of
the unit cell are taken at random and checked to determine whether the two
surfaces agree, or not, in the assignment of the point to the A-rich region.
The relevant quantity is $(\delta V)\Delta/VN^{1/2}a$, the fraction of the
volume $\delta V$ for which there is disagreement, normalized by the area of
the grain boundary $V/\Delta$. The factor $N^{1/2}a$ ensures that this
measure, which we denote $I'_{\mathrm{S}}$, is dimensionless.

\begin{figure}

\includegraphics[width=75mm]{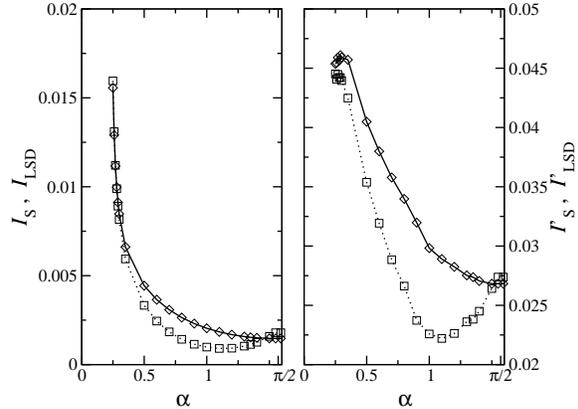}
\caption{%
Measures of the similarity between the intermaterial
dividing surface defined by our calculation and that of Scherck's first
surface (diamonds, joined by solid line) and that defined by
the linear stack of dislocations
(squares, dotted line). Except near twist angles of $\pi/2$,
the latter is a better representation.}
\end{figure}

These two measures are plotted in Fig. 5. They indicate that the LSD
is a better representation of the intermaterial dividing surface
over almost the entire range of twist angles except, perhaps, quite close to
$\pi/2$. As the angle decreases, both measures tend to the same limit, of
course. (Recall the only difference between the two surfaces is a dilation
of $\cos(\alpha/2)$.)

\begin{figure}

\includegraphics[width=75mm]{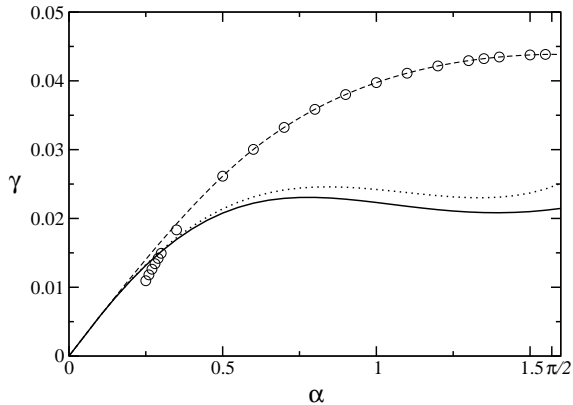}
\caption{%
Free energy per unit area, $\gamma$, of the twist grain
boundary as a function of twist angle $\alpha$. The circles show our
results, the dashed line is a fit to them (extrapolated to $\alpha=0$)
taking into account only data for $\alpha>0.5.$ Also shown are the
approximate free energies we have calculated from the expressions in Ref.
\cite{k&l} for Scherck's first surface (solid line) and the linear array of
dislocations (dotted line) which include only bending and compression
energies.}

\end{figure}

The grain boundary free energy as a function of the twist angle is shown in
Fig. 6. The circles show the results of our calculation. For values of the
twist angle greater than $0.5$ radians, our results are accurate to within
$5\%$, improving towards $1\%$ as they approach $\pi/2$.
The dashed line has been drawn through these values and extrapolated
to zero. We have also included several other points for smaller
values of $\alpha$. The number of basis functions employed is insufficient
for the grain boundary free energy to have converged to within $5\%$.
Nevertheless, we have included them as they appear to indicate that the
behavior of the free energy as the angle approaches zero may not be linear,
the behavior expected if, at very low densities, dislocations repel one
another\cite{k&l}. It has been argued, however, that at very low densities,
dislocations attract one another\cite{luk}.

One notes that the grain boundary free energies are rather small: that of
the grain boundary with twist angle of $\pi/2$ is, at the same
incompatibility, $\chi N$, somewhat less than half the energy of the
boundary with tilt angle of $\pi/2$\cite{matsen}. Perhaps this should not be
too surprising. In the approximation noted earlier of treating the
intermaterial dividing surface as a surface of constant mean curvature, with
an energy given by the Helfrich free energy\cite{wands}, the grain
boundary free energy would be identically zero\cite{gg}.

The free energy of a system with a twist grain boundary was calculated
previously by Gido and Thomas\cite{gandt}. They applied a version of the
self consistent field to a brush of infinitely stretched chains anchored to
a given saddle-shaped surface, and also carried out an independent
calculation based on work of Wang and Safran\cite{wands}. However they
report their results in terms of the extensive free energy per chain in the
region of Scherck's surface. This is not a uniquely defined quantity, nor is
it the thermodynamic grain boundary free energy per unit area which we have
calculated, so direct comparisons are precluded.

We have chosen to compare our results with those of Kamien and
Lubensky\cite{k&l}. We emphasize that the two calculations are rather
different in principle. In the approach we have employed, the free energy of
the block copolymer system is calculated directly, assuming nothing other
than the applicability of self consistent field theory. In particular, we do
not employ elasticity theory, or assume that displacements from a reference
system without a grain boundary are small, \textit{etc}.
In contrast, that of Ref.
\cite{k&l} assumes that the bulk system can be adequately described as a
series of surfaces, and the energy of this system of surfaces can be
expanded assuming small displacements. A further difficulty which arises
when applying the calculation of  Ref. \cite{k&l} to a block copolymer
system is that the volumes on either side of the surfaces  are
undifferentiated, whereas in the block copolymer system, these volumes  are
filled with \emph{different} monomers. Thus the symmetry of the system
considered by Kamien and Lubensky is not the same as that of ours. As a
consequence, there are more elastic constants in an elastic description of a
block copolymer lamellar phase than the two utilized by them in their
description of liquid crystalline smectics\cite{landau}.

Having acknowledged these \textit{caveats}, we calculate the bending and
compression contributions to the free energies of Scherck's first surface
and of an LSD surface as given in  Ref. \cite{k&l}. As noted in the
Introduction, a number of parameters needed to evaluate these free energies
are unknown, being inputs to the phenomenological theory. However we can
provide some of these values from our work.  Thus we calculate the lamellar
spacing to be $D=1.5155\ aN^{1/2}$, and the dimensionless compression
modulus to be $B=3.01$. The dimensionless  bending modulus is unknown but
can be estimated \cite{kappa} to be $\kappa\approx 0.115$. The only unknown
which remains is the size of the ``core region", which provides a cutoff to
the otherwise divergent integrals for the compression free energy estimated
in \cite{k&l}. This can be estimated from the slope of the dotted line in
Fig. 6. We obtain, thereby, a value of $\approx 0.191 a N^{1/2}$,
\textit{i.e.}, $25\%$ of the IMDS spacing $D/2$. Of course, there is no
reason that this core region should not depend on $\alpha$, but were we to
obtain the size of the core from our data at each value of the twist angle,
the result would simply be a mapping of the results of Ref. \cite{k&l} to
ours, and no independent comparison would be possible. Using these
parameters, we have evaluated the free energy given in  Ref. \cite{k&l} of
the appropriate Scherck's surface and of the linear superposition of
dislocations. These are shown in Fig. 6 as solid and dotted lines
respectively. The LSD has a  free energy closer to our result than does
Scherck's surface, just as it is closer to our intermaterial dividing
surface. Both approximations underestimate the grain boundary free energy of
the block copolymer system by a factor which increases with twist angle, and
is about two at $\alpha=\pi/2$.

\section{Conclusions and outlook}

We have applied self consistent field theory to twist grain boundaries in
block copolymer melts. Our calculation is more direct than earlier ones and
provides greater information concerning the monomer densities throughout the
volume. It also expresses the grain boundary free energy in terms of the
directly measurable volume per chain and radius of gyration as opposed to
elastic modulii of internal interfaces. The boundary free energy was
obtained as a function of twist angle, and found to be quite small; smaller
than kink grain boundaries of the same angle and incompatibility.

We have compared our results to previous phenomenological calculations to
show that the intermaterial dividing surface is not given by either surface,
Scherck's first surface or the linear stack of dislocations, but that of the
two, the latter is a better representation over most twist angles except
near $\pi/2$.

We comment briefly on the other type of twist grain boundary which has been
observed at small twist angles\cite{gidotgb} , the one consisting on a stack
of lamellae which are twisted slightly and remain continuous. We have not
investigated it is because we failed to find an appropriate periodic
boundary condition which does not contribute to the excess surface free
energy. Although it is possible to calculate the contribution to the excess
surface free energy of any choice of boundary condition and then to subtract
it from the total excess, leaving the desired grain boundary free energy,
the procedure is tedious.  However simple examination of this boundary leads
to the conclusion that the grain boundary free energy must be approximately
twice that of a kink grain boundary. This is because the lamellae within the
boundary and  those far from it meet in what approximates a kink grain
boundary. (Fig. 4 of Ref. \cite{gidotgb} shows this nicely.) The free energy
of a kink grain boundary grows as $\theta^3$ for small kink angle
$\theta$\cite{nas,matsen}. Of course we do not know the relation between the
angle, $\theta$, of this ``effective" kink grain boundary and the twist
angle $\alpha$. Nonetheless, if we assume that the relation is linear, then
the twist grain boundary energy would grow as $\alpha^3$ for small twist
angles and would be favored over those we have modeled here, which would in
fact be metastable but long-lived as their energy is small. This is in
accord with the experimental result that both forms of boundary are observed
at small twist angles. But at larger angles such grain boundaries would be
disfavored compared to those considered in this paper. This is in accord
with the fact that they are not seen experimentally.

\section*{Acknowledgments}

We  are grateful to  Xiao-Jun Li for his comments and assistance, and thank
David Andelman and Yoav Tsori for correspondence. One of us, (M.S.), would
like to thank Holm Holmsen, Aurora Martinez, and the Department of
Biochemistry and Molecular Biology of the University of Bergen for their
hospitality while this paper was written. This work was supported in part by
grants from the United States-Israel Binational Science Foundation under
grant 98-00429, and the National Science Foundation, under grant number
DMR9876864.


\begin{thebibliography}{99}
\bibitem{gg} We are indebted to Gerhard Gompper for reminding us of this.
\bibitem{kappa} Matsen \cite{matsen} shows
how $\kappa$ can be determined from an analysis of the kink grain boundary
at low angles. His analysis for $\chi N=20$ yields $\kappa=0.17$. We could
repeat his calculations for our value of $\chi N=15$, but as this modulus
scales with the energy,  we can approximate the modulus at our value of
incompatibility by  $\kappa\approx 0.17\times 15/20=0.13$. Note that the
same scaling of his result for the compression modulus $B=4.17$ at $\chi
N=20$  yields $B\approx 3.12$ at $\chi N=15$, which is very close to the
value of $B=3.01$ which we calculate directly.
\end{thebibliography}
\end{document}